\documentclass[english]{article}%
\usepackage{amssymb}
\usepackage{fontenc}
\usepackage[latin1]{inputenc}
\usepackage{amsmath}
\usepackage{fontenc}
\usepackage{babel}
\usepackage{amsfonts}
\usepackage{graphicx}%
 
\begin{document}

\title{{\Large Coherent and semiclassical states of a free particle}}
\author{V.G. Bagrov\thanks{Department of Physics, Tomsk State University, Russia;
e-mail: bagrov@phys.tsu.ru}, D. M. Gitman\thanks{P.N. Lebedev Physical
Institute, Moscow, Russia; Tomsk State University, Russia; Institute of
Physics, University of São Paulo, Brazil; e-mail: gitman@dfn.if.usp.br. }, and
A. S. Pereira\thanks{Institute of Physics, University of São Paulo, Brazil;
e-mail: apereira@if.usp.br}}
\maketitle

\begin{abstract}
Coherent states (CS) were first introduced and studied in detail for bound
motion and discrete-spectrum systems like harmonic oscillators and similar
systems with a quadratic Hamiltonian. However, the problem of constructing CS
has still not been investigated in detail for the simplest and physically
important case of a free particle, for which, besides being physically
important, the CS problem is of didactic value in teaching quantum mechanics,
with the CS regarded as examples of wave packets representing semiclassical
motion. In this paper, we essentially follow the Malkin-Dodonov-Man'ko method
to construct the CS of a free nonrelativistic particle. We give a detailed
discussion of the properties of the CS obtained, in particular, the
completeness relations, the minimization of uncertainty relations, and the
evolution of the corresponding probability density. We describe the physical
conditions under which free-particle CS can be considered semiclassical states.


\end{abstract}

\section{Introduction}

Coherent states (CS) play an important role in modern quantum theory as states
that provide a natural relation between quantum mechanical and classical
descriptions. They have a number of useful properties and, as a consequence a
wide range of applications, e.g., in semiclassical description of quantum
systems, in quantization theory, in condensed matter physics, in radiation
theory, in quantum computations (see, e.g., Refs. \cite{CSQT72,MalMa,DodMa87}%
). Although there are numerous publications devoted to constructing CS of
different systems, a universal definition of a CS and a workable scheme to
construct them for an arbitrary physical system is not known. However, we
believe that the problem of constructing CS for systems with quadratic
Hamiltonians of the general form was completely solved in works by Dodonov and
Man'ko, using Malkin and Man'ko integral of motion method, (see
\cite{MalMa,DodMa87,DodMa03}). It should be noted that extracting concrete
sets of CS and their properties (for a given quadratic system) from their
general results one has to perform an additional technical efforts. In this
article, we turn our attention to the CS of a free particle. Besides of their
physical importance there is a didactic advantage of using free-particle CS in
teaching of quantum mechanics, considering them as examples of exact wave
packages representing the semiclassical particle motion. In this relation, we
note that CS were first introduced and studied in detail for systems with
bounded motion and discrete spectrum like harmonic oscillator or a charged
particle in a magnetic field. However, for such a simple and physically
important system as a free particle, the problem of CS construction was not
solved that time. We believe that this situation is explained by the fact that
the free particle represents an unbounded motion with the continuous energy
spectrum, and a generalization of the initial (Glauber) scheme in constructing
CS of a harmonic oscillator was not so obvious in this case. Although CS of a
free particle, in principle, could be extracted from the above mentioned
general results of Dodonov and Man'ko, many authors (ignoring or simply
unaware of their results) keep trying to construct CS of a free particle,
inventing their own ways. Describing these attempts, we have to cite Refs.
\cite{Littl86,TorGo10,GueLoA11,GelHnK12} devoted to this problem. In our
opinion, no single one of these studies completely solves the problem under
consideration. Authors of Ref. \cite{TorGo10} have quite closely approached
the goal, choosing a particular case of initial states for their CS.\ But even
for such initial states, they did not derive an explicit form of time
dependent free particle CS and did not study their properties. In fact, their
program was realized in the work \cite{Littl86}, but the author did not
identify his states with some kind of CS. In \cite{GueLoA11}, the authors
consider the limit of zero frequency in CS of a harmonic oscillator, deriving
a sort of CS for a free particle. Their CS are expressed in terms of sums of
Hermite polynomials and the complicated form of the CS hampers their
interpretation, study, and applications. Another study \cite{GelHnK12} treats
free particle CS in the framework of a general approach to constructing CS for
system with a continuous spectrum. The approach is based on using
nonnormalizable fiducial states and involves quite complicated techniques. The
authors do not present free-particle well defined for any time instants.

In the present article, in facto following the Dodonov-Man'ko method, we
construct different families of generalized CS of a free massive
nonrelativistic particle. We discuss in detail properties of the constructed
CS, in particular, completeness relations, minimization of uncertainty
relations and evolution of the corresponding probability density in time. We
describe physical conditions when free particle CS can be considered as
semiclassical states.

\section{Constructing time-dependent CS of a free particle}

\subsection{Basic equations}

For simplicity, we consider one-dimensional quantum motion of a free
nonrelativistic particle of the mass $m$ on the whole real axis $\mathbb{R}%
=\left(  -\infty,\infty\right)  .$ It is described by the Schrödinger
equation
\begin{equation}
i\hbar\partial_{t}\Psi\left(  x,t\right)  =\hat{H}_{x}\Psi\left(  x,t\right)
,\ x\in\mathbb{R},\label{b1}%
\end{equation}
where the Hamiltonian $\hat{H}_{x}$ and the momentum operator $\hat{p}_{x}$,
\begin{equation}
\hat{H}_{x}=-\frac{\hbar^{2}}{2m}\partial_{x}^{2}=\frac{\hat{p}_{x}^{2}}%
{2m},\ \hat{p}_{x}=-i\hbar\partial_{x},\label{b1a}%
\end{equation}
are self-adjoint on their natural domains, \cite{book2}.

It is useful to introduce the dimensionless variables%
\begin{equation}
q=xl^{-1},\text{ \ }\tau=\frac{\hbar}{ml^{2}}t.\label{b2}%
\end{equation}
Then eq. (\ref{b1}) takes the form%
\begin{align}
&  \hat{S}\psi\left(  q,\tau\right)  =0,\text{ \ }\hat{S}=i\partial_{\tau
}-\hat{H},\ \hat{H}_{x}=\frac{\hbar^{2}}{ml^{2}}\hat{H}\ ,\nonumber\\
&  \hat{H}=\frac{\hat{p}^{2}}{2},\ \hat{p}=-i\partial_{q},\ \psi\left(
q,\tau\right)  =\sqrt{l}\Psi\left(  lq,\frac{ml^{2}}{\hbar}\tau\right)
,\label{b4}%
\end{align}
with $\left\vert \Psi\left(  x,t\right)  \right\vert ^{2}dx=\left\vert
\psi\left(  q,\tau\right)  \right\vert ^{2}dq$. We call the operator $\hat{S}$
the equation operator.

In terms of creation and annihilation operators $\hat{a}$ and $\hat{a}^{\dag}%
$,%
\begin{equation}
\hat{a}=\frac{\hat{q}+i\hat{p}}{\sqrt{2}},\text{ \ }\hat{a}^{\dag}=\frac
{\hat{q}-i\hat{p}}{\sqrt{2}},\ \left[  \hat{a},\hat{a}^{\dag}\right]
=1,\nonumber
\end{equation}
the Hamiltonian $\hat{H}$ is a quadratic form of these operators
\begin{equation}
\hat{H}=\frac{1}{4}\left[  \hat{a}^{\dag}\hat{a}+\hat{a}\hat{a}^{\dag}-\left(
\hat{a}^{\dag}\right)  ^{2}-\hat{a}^{2}\right]  . \label{4.1}%
\end{equation}
It cannot be reduced to the first canonical form for a quadratic combination
of creation and annihilation operators, which is the oscillator-like form, by
any canonical transformation; this indicates that the spectrum of $\hat{H}$ is
continuous, see eg. \cite{105}.

\subsection{Integrals of motion linear in canonical operators $\hat{q}$ and
$\hat{p}$}

We construct an integral of motion $\hat{A}\left(  \tau\right)  $ linear in
$\hat{q}$ and $\hat{p}$. The general form of such an integral of motion reads
\begin{equation}
\hat{A}\left(  \tau\right)  =f\left(  \tau\right)  \hat{q}+ig\left(
\tau\right)  \hat{p}+\varphi\left(  \tau\right)  ,\label{3.1}%
\end{equation}
where $f\left(  \tau\right)  $, $g\left(  \tau\right)  $ and $\varphi\left(
\tau\right)  $ are some complex functions of the time $\tau$. For the operator
$\hat{A}\left(  \tau\right)  $ to be an integral of motion, it has to commute
with the equation operator (\ref{b4}),
\begin{equation}
\left[  \hat{S},\hat{A}\left(  \tau\right)  \right]  =0.\label{3.2}%
\end{equation}
If the Hamiltonian is self-adjoint, the adjoint operator $\hat{A}^{\dag
}\left(  \tau\right)  $ is also an integral of motion, $\left[  \hat{S}%
,\hat{A}^{\dag}\left(  \tau\right)  \right]  =0.$

Substituting representations (\ref{3.1}) into eqs. (\ref{3.2}), we obtain the
following equations for the functions $f\left(  \tau\right)  $, $g\left(
\tau\right)  ,$ and $\varphi\left(  \tau\right)  $:
\begin{equation}
\dot{f}\left(  \tau\right)  =0,\ \dot{g}\left(  \tau\right)  -if\left(
\tau\right)  =0,\ \ \dot{\varphi}\left(  \tau\right)  =0,\label{3.3}%
\end{equation}
where the dot denote derivatives with respect to $\tau$. The general solution
of eqs. (\ref{3.3}) is
\begin{equation}
f\left(  \tau\right)  =c_{1},\ g\left(  \tau\right)  =c_{2}+ic_{1}%
\tau,\ \ \varphi\left(  \tau\right)  =\mathrm{const},\label{3.4}%
\end{equation}
where $c_{1}$ and $c_{2}$ are arbitrary constants. Without loss of the
generality we can set $\varphi\left(  \tau\right)  =0.$ Thus,%
\begin{equation}
\hat{A}\left(  \tau\right)  =c_{1}\hat{q}+ig\left(  \tau\right)  \hat
{p},\ \ g\left(  \tau\right)  =c_{2}+ic_{1}\tau.\label{3.4a}%
\end{equation}

The commutator $\left[  \hat{A}\left(  \tau\right)  ,\hat{A}^{\dag}\left(
\tau\right)  \right]  $ reads
\begin{equation}
\left[  \hat{A}\left(  \tau\right)  ,\hat{A}^{\dag}\left(  \tau\right)
\right]  =2\operatorname{Re}\left(  g^{\ast}\left(  \tau\right)  f\left(
\tau\right)  \right)  =2\operatorname{Re}\left(  c_{1}^{\ast}c_{2}\right)
=\delta. \label{3.5}%
\end{equation}

Equations (\ref{3.4}) imply that $\delta$ is real-valued integral of motion,
$\delta=\mathrm{const}$. In what follows we set $\delta=1$,%
\begin{equation}
\delta=2\operatorname{Re}\left(  c_{1}^{\ast}c_{2}\right)  =1.\label{3.7}%
\end{equation}
Let $c_{1}=\left\vert c_{1}\right\vert e^{i\mu_{1}}$ and $c_{2}=\left\vert
c_{2}\right\vert e^{i\mu_{2}}$. Condition (\ref{3.7}) then implies that%
\begin{equation}
\left\vert c_{2}\right\vert \left\vert c_{1}\right\vert \cos\left(  \mu
_{2}-\mu_{1}\right)  =\frac{1}{2}.\label{3.7b}%
\end{equation}
Choosing $\delta=1$, we set $\hat{A}\left(  \tau\right)  $ and $\hat{A}^{\dag
}\left(  \tau\right)  $ to be annihilation and creation operators,%
\begin{equation}
\left[  \hat{A}\left(  \tau\right)  ,\hat{A}^{\dag}\left(  \tau\right)
\right]  =1.\label{3.7a}%
\end{equation}

It follows from eqs. (\ref{3.4a}) and (\ref{3.7}) that%
\begin{align}
&  \hat{q}=g^{\ast}\left(  \tau\right)  \hat{A}\left(  \tau\right)  +g\left(
\tau\right)  \hat{A}^{\dag}\left(  \tau\right)  ,\ \ g\left(  \tau\right)
=c_{2}+ic_{1}\tau,\nonumber\\
&  i\hat{p}=c_{1}^{\ast}\hat{A}\left(  \tau\right)  -c_{1}\hat{A}^{\dag
}\left(  \tau\right)  .\label{3.8}%
\end{align}

We note that the operators $\hat{q}$ and $\hat{p}$ cannot depend on the
constants $c_{1}$, $c_{2}$ and time $\tau$. Indeed, using eqs. (\ref{3.1}) and
(\ref{3.7}), one can verify that relations $\partial_{\tau}\hat{q}%
=\partial_{\tau}\hat{p}=\partial_{c_{1}}\hat{p}=\partial_{c_{1}}\hat
{q}=\partial_{c_{2}}\hat{p}=\partial_{c_{2}}\hat{q}=0$ hold true.

\subsection{Time-dependent generalized CS}

We consider eigenvectors $\left\vert z,\tau\right\rangle $ of the annihilation
operator $\hat{A}\left(  \tau\right)  $ corresponding to the eigenvalue $z$%
\begin{equation}
\hat{A}\left(  \tau\right)  \left\vert z,\tau\right\rangle =z\left\vert
z,\tau\right\rangle .\label{4.1a}%
\end{equation}
In general, $z$ is a complex number. It follows from eqs. (\ref{3.8}) and
(\ref{4.1a}) that%
\begin{align}
&  q\left(  \tau\right)  \equiv\left\langle z,\tau\left\vert \hat
{q}\right\vert z,\tau\right\rangle =q_{0}+p\tau,\ q_{0}=c_{2}^{\ast}%
z+c_{2}z^{\ast},\nonumber\\
&  p\left(  \tau\right)  \equiv\left\langle z,\tau\left\vert \hat
{p}\right\vert z,\tau\right\rangle =i\left(  c_{1}z^{\ast}-c_{1}^{\ast
}z\right)  =p,\nonumber\\
&  z=c_{1}q\left(  \tau\right)  +ig\left(  \tau\right)  p=c_{1}q_{0}%
+ic_{2}p.\label{4.3}%
\end{align}
Written in $q$-representation, eqn. (\ref{4.1a}) becomes%
\begin{equation}
\left[  c_{1}q+g\left(  \tau\right)  \partial_{q}\right]  \Phi_{z}^{c_{1,2}%
}\left(  q,\tau\right)  =z\Phi_{z}^{c_{1,2}}\left(  q,\tau\right)
,\ \ \Phi_{z}^{c_{1,2}}\left(  q,\tau\right)  \equiv\left\langle
q|z,\tau\right\rangle .\label{c2}%
\end{equation}

The general solution of this equation has the form%
\begin{equation}
\langle q\left\vert z,\tau\right\rangle =\Phi_{z}^{c_{1,2}}\left(
q,\tau\right)  =\exp\left[  -\frac{c_{1}}{g\left(  \tau\right)  }\frac{q^{2}%
}{2}+\frac{zq}{g\left(  \tau\right)  }+\chi\left(  \tau,z\right)  \right]
,\label{4.5}%
\end{equation}
where $\chi\left(  \tau,z\right)  $ is an arbitrary function on $\tau$ and
$z.$

We can see that the functions $\Phi_{z}^{c_{1,2}}\left(  q,\tau\right)  $ can
be written in terms of the mean values $q\left(  \tau\right)  $ and $p\left(
\tau\right)  $,%
\begin{equation}
\Phi_{z}^{c_{1,2}}\left(  q,\tau\right)  =\exp\left\{  ipq-\frac{c_{1}%
}{2g\left(  \tau\right)  }\left[  q-q\left(  \tau\right)  \right]  ^{2}%
+\phi\left(  \tau,z\right)  \right\}  .\label{4.6a}%
\end{equation}
where $\phi\left(  \tau,z\right)  $ is again an arbitrary function on $\tau$
and $z$.

The functions $\Phi_{z}$ satisfy the following equation%
\begin{equation}
\hat{S}\Phi_{z}^{c_{1,2}}\left(  q,\tau\right)  =\lambda\left(  \tau,z\right)
\Phi_{z}^{c_{1,2}}\left(  q,\tau\right)  ,\label{4.7}%
\end{equation}
where%
\begin{equation}
\lambda\left(  \tau,z\right)  =i\dot{\phi}\left(  \tau,z\right)  -\frac{1}%
{2}\left[  p^{2}+\frac{c_{1}}{g\left(  \tau\right)  }\right]  .\label{4.9}%
\end{equation}
For functions (\ref{4.6a}) to satisfy Schrödinger equation (\ref{b4}), we have
to fix $\phi\left(  \tau,z\right)  $ from the condition $\lambda\left(
\tau,z\right)  =0$. Thus, for the function $\phi\left(  \tau,z\right)  $, we
obtain%
\begin{equation}
\phi\left(  \tau,z\right)  =-\frac{i}{2}p^{2}\tau-\frac{1}{2}\ln g\left(
\tau\right)  +\ln N,\label{4.12}%
\end{equation}
where $N$ is a normalization constant, which we suppose to be real.

The density probability generated by function (\ref{4.6a}) reads%
\begin{equation}
\rho\left(  q,\tau\right)  =\left\vert \Phi_{z}^{c_{1,2}}\left(
q,\tau\right)  \right\vert ^{2}=\frac{N^{2}}{\left\vert g\left(  \tau\right)
\right\vert }\exp\left\{  -\frac{\left[  q-q\left(  \tau\right)  \right]
^{2}}{2\left\vert g\left(  \tau\right)  \right\vert ^{2}}\right\}  .
\label{4.15}%
\end{equation}
Considering the normalization integral, we find the constant $N$,%
\begin{equation}
\int_{-\infty}^{\infty}\rho\left(  q,\tau\right)  dq=1\Rightarrow N=\left(
2\pi\right)  ^{-1/4}. \label{4.16}%
\end{equation}
Thus, normalized solutions of the Schrödinger equation that are eigenfunctions
of the annihilation operator $\hat{A}\left(  \tau\right)  $ have the form%
\begin{equation}
\Phi_{z}^{c_{1,2}}\left(  q,\tau\right)  =\frac{1}{\sqrt{\sqrt{2\pi}g\left(
\tau\right)  }}\exp\left\{  i\left(  pq-\frac{1}{2}p^{2}\tau\right)
-\frac{c_{1}}{g\left(  \tau\right)  }\frac{\left[  q-q\left(  \tau\right)
\right]  ^{2}}{2}\right\}  \label{4.13}%
\end{equation}
and the corresponding probability density reads%
\begin{equation}
\rho_{z}^{c_{1,2}}\left(  q,\tau\right)  =\left\vert \Phi_{z}^{c_{1,2}}\left(
q,\tau\right)  \right\vert ^{2}=\frac{1}{\sqrt{2\pi}\left\vert g\left(
\tau\right)  \right\vert }\exp\left\{  -\frac{\left[  q-q\left(  \tau\right)
\right]  ^{2}}{2\left\vert g\left(  \tau\right)  \right\vert ^{2}}\right\}  .
\label{4.14}%
\end{equation}

In what follows, we call the solutions (\ref{4.13}) the time-dependent
generalized CS. In fact, we have a family of states parametrized by two
complex constants $c_{1}$ and $c_{2}$ that satisfy restriction (\ref{3.7}). As
we see in what follows, each family of the generalized CS represent so-called
squeezed states. Additional restrictions on the constants $c_{1}$ and $c_{2}$
transform these states into CS of the free particle (see below).

We note that the generalized CS can be constructed in the Glauber manner,
acting by the displacement operator $\mathcal{D}\left(  z,\tau\right)
=\exp\left[  z\hat{A}^{\dag}\left(  \tau\right)  -z^{\ast}\hat{A}\left(
\tau\right)  \right]  $ on the vacuum vector $\left\vert 0,\tau\right\rangle $
defined as $\hat{A}\left(  \tau\right)  \left\vert 0,\tau\right\rangle =0$:%
\begin{align}
&  \left\langle q|z,\tau\right\rangle =\mathcal{D}\left(  z,\tau\right)
\langle q\left\vert 0,\tau\right\rangle =\exp\left[  -\frac{\left\vert
z\right\vert ^{2}}{2}\right]  \sum_{n=0}^{\infty}\frac{z^{n}}{\sqrt{n!}%
}\langle q\left\vert n,\tau\right\rangle ,\nonumber\\
&  \left\vert n,\tau\right\rangle =\frac{\left[  \hat{A}^{\dag}\left(
\tau\right)  \right]  ^{n}}{\sqrt{n!}}\left\vert 0,\tau\right\rangle
,\ \left\vert 0,\tau\right\rangle =\frac{1}{\sqrt{\sqrt{2\pi}g\left(
\tau\right)  }}\exp\left\{  -\frac{c_{1}}{g\left(  \tau\right)  }\frac{q^{2}%
}{2}\right\}  .\label{w1}%
\end{align}
Functions (\ref{w1}) differ from the set (\ref{4.13}) by a constant phase
factor only.

Using completeness property of the states $\left\vert n,\tau\right\rangle ,$%
\begin{equation}
\sum_{n=0}^{\infty}\left\vert n,\tau\right\rangle \left\langle n,\tau
\right\vert =1,\ \ \forall\tau,\label{9}%
\end{equation}
we can find the overlapping and prove the completeness relations for the
generalized CS of the free particle%
\begin{align}
&  \left\langle z^{\prime},\tau|z,\tau\right\rangle =\exp\left(  z^{\prime
\ast}z-\frac{\left\vert z^{\prime}\right\vert ^{2}+\left\vert z\right\vert
^{2}}{2}\right)  ,\ \ \forall\tau;\nonumber\\
&  \int\int\left\langle q|z,\tau\right\rangle \left\langle z,\tau|q^{\prime
}\right\rangle d^{2}z=\pi\delta\left(  q-q^{\prime}\right)  ,\text{ \ }%
d^{2}z=d\operatorname{Re}zd\operatorname{Im}z,\text{ \ }\forall\tau.\label{10}%
\end{align}

\section{Standard deviations, uncertainty relations, and CS of a free
particle}

Calculating standard deviations $\sigma_{q}\left(  \tau\right)  $, $\sigma
_{p}$, and the quantity $\sigma_{qp}\left(  \tau\right)  $ in the generalized
CS, we obtain%
\begin{align}
&  \sigma_{q}\left(  \tau\right)  =\sqrt{\langle\left(  \hat{q}-\left\langle
q\right\rangle \right)  ^{2}\rangle}=\sqrt{\left\langle q^{2}\right\rangle
-\left\langle q\right\rangle ^{2}}=\left\vert g\left(  \tau\right)
\right\vert ,\nonumber\\
&  \sigma_{p}\left(  \tau\right)  =\sqrt{\langle\left(  \hat{p}-\left\langle
p\right\rangle \right)  ^{2}\rangle}=\sqrt{\left\langle p^{2}\right\rangle
-\left\langle p\right\rangle ^{2}}=\left\vert f\left(  \tau\right)
\right\vert =\left\vert c_{1}\right\vert ,\nonumber\\
&  \sigma_{qp}\left(  \tau\right)  =\frac{1}{2}\left\langle \left(  \hat
{q}-\left\langle q\right\rangle \right)  \left(  \hat{p}-\left\langle
p\right\rangle \right)  +\left(  \hat{p}-\left\langle p\right\rangle \right)
\left(  \hat{q}-\left\langle q\right\rangle \right)  \right\rangle \nonumber\\
&  =i\left[  1/2-g\left(  \tau\right)  f^{\ast}\left(  \tau\right)  \right]
.\label{d3}%
\end{align}
It is easy to see that the generalized CS minimize the Robertson-Schrödinger
uncertainty relation \cite{SchroRo},%
\begin{equation}
\sigma_{q}^{2}\left(  \tau\right)  \sigma_{p}^{2}-\sigma_{qp}^{2}\left(
\tau\right)  =\frac{1}{4}.\label{d.3}%
\end{equation}
This means that these states are squeezed states for any time instant.

We evaluate the Heisenberg uncertainty relation in the generalized CS taking
constraint (\ref{3.7}) into account:%
\begin{equation}
\left.  \sigma_{q}\left(  \tau\right)  \sigma_{p}\left(  \tau\right)
\right\vert _{2\operatorname{Re}\left(  c_{1}^{\ast}c_{2}\right)  }%
=\sqrt{\frac{1}{4}+\left[  \left\vert c_{2}\right\vert \left\vert
c_{1}\right\vert \sin\left(  \mu_{2}-\mu_{1}\right)  +\left\vert
c_{1}\right\vert ^{2}\tau\right]  ^{2}}\geq\frac{1}{2}.\label{5.5}%
\end{equation}
Using (\ref{d3}), we then find $\sigma_{q}\left(  0\right)  =\sigma
_{q}=\left\vert c_{2}\right\vert $ and $\sigma_{p}\left(  0\right)
=\sigma_{p}=\left\vert c_{1}\right\vert $, and hence at $\tau=0$ this relation
become
\begin{equation}
\left.  \sigma_{q}\sigma_{p}\right\vert _{2\operatorname{Re}\left(
c_{1}^{\ast}c_{2}\right)  }=\left\vert c_{2}\right\vert \left\vert
c_{1}\right\vert =\sqrt{\frac{1}{4}+\left[  \left\vert c_{2}\right\vert
\left\vert c_{1}\right\vert \sin\left(  \mu_{2}-\mu_{1}\right)  \right]  ^{2}%
}.\label{F4a}%
\end{equation}
We see that $\left\vert c_{i}\right\vert \neq0$, $i=1,2$ and the left-hand
side of (\ref{F4a}) is minimal if $\mu_{1}=\mu_{2}=\mu,$ which provides the
minimization of the Heisenberg uncertainty relation in the generalized CS at
the initial time instant,%
\begin{equation}
\sigma_{q}\sigma_{p}=\frac{1}{2}.\label{F5}%
\end{equation}

In what follows, we consider the free-particle generalized CS with the
restriction $\mu_{1}=\mu_{2}$. Such states are simply called CS of a free particle.

Now, constraint (\ref{3.7}) take the form%
\begin{equation}
\left\vert c_{2}\right\vert \left\vert c_{1}\right\vert =1/2\Longrightarrow
c_{2}^{\ast}=c_{1}^{-1}/2.\label{F6}%
\end{equation}
We see that the constant $\mu$ don't enter the CS (\ref{4.13}). Thus, we set
$\mu=0$ in what follows. Then%
\begin{align}
&  c_{2}=\left\vert c_{2}\right\vert =\sigma_{q},\ c_{1}=\left\vert
c_{1}\right\vert =\sigma_{p}=1/(2\sigma_{q}),\nonumber\\
&  g\left(  \tau\right)  =\left(  \sigma_{q}+\frac{i\tau}{2\sigma_{q}}\right)
,\ \sigma_{q}\left(  \tau\right)  =\left\vert g\left(  \tau\right)
\right\vert =\sqrt{\sigma_{q}^{2}+\frac{\tau^{2}}{4\sigma_{q}^{2}}%
}.\label{d12}%
\end{align}

From eqn. (\ref{d12}), we conclude that for any $\tau$, the Heisenberg
uncertainty relation in the CS takes the form
\begin{equation}
\sigma_{q}\left(  \tau\right)  \sigma_{p}=\frac{1}{2}\sqrt{1+\frac{\tau^{2}%
}{4\sigma_{q}^{4}}}\geq\frac{1}{2}\label{d13}%
\end{equation}
and CS of a free particle read%
\begin{equation}
\Phi_{z}^{\sigma_{q}}\left(  q,\tau\right)  =\frac{\exp\left\{  i\left(
pq-\frac{p^{2}}{2}\tau\right)  -\frac{\left[  q-q\left(  \tau\right)  \right]
^{2}}{4\left(  \sigma_{q}^{2}+i\tau/2\right)  }\right\}  }{\sqrt{\left(
\sigma_{q}+\frac{i\tau}{2\sigma_{q}}\right)  \sqrt{2\pi}}}.\label{d14}%
\end{equation}

In fact, we have a family of CS parametrized by one real parameter $\sigma
_{q}$. Each set of CS in the family has its specific initial standard
deviations $\sigma_{q}>0$. CS from a family with a given $\sigma_{q}$ are
labeled by quantum numbers $z$,%
\begin{equation}
z=\frac{q_{0}}{2\sigma_{q}}+i\sigma_{q}p,\label{d14a}%
\end{equation}
which are in one to one correspondence with trajectory initial data $q_{0}$
and $p,$%
\begin{equation}
q_{0}=2\sigma_{q}\operatorname{Re}z\ ,\ \ p=\frac{\operatorname{Im}z}%
{\sigma_{q}}\ .\label{d15a}%
\end{equation}

The probability densities that corresponds to the CS are
\begin{equation}
\rho_{z}^{\sigma_{q}}\left(  q,\tau\right)  =\frac{1}{\sqrt{\left(  \sigma
_{q}^{2}+\frac{\tau^{2}}{4\sigma_{q}^{2}}\right)  2\pi}}\exp\left\{  -\frac
{1}{2}\frac{\left[  q-q\left(  \tau\right)  \right]  ^{2}}{\sigma_{q}%
^{2}+\frac{\tau^{2}}{4\sigma_{q}^{2}}}\right\}  . \label{d15}%
\end{equation}

It can be seen that at any time instant $\tau$, \ probability densities
(\ref{d15}) are given by Gaussian distributions with standard deviations
$\sigma_{q}\left(  \tau\right)  .$ The mean values $\left\langle
q\right\rangle =q\left(  \tau\right)  =q_{0}+p\tau$ are moving along the
classical trajectory with the particle velocity $p$. The maxima of the
probability densities move with the same velocity (\ref{d15}).

Figure 1 plots function (\ref{d15}) with $\sigma=2^{-1/2}$, $p=2,$ $q_{0}=0,$
for two time instants, $\tau=0$ and $\tau=1$,%

\begin{figure}[th]
\begin{center}
\includegraphics[scale=0.5]{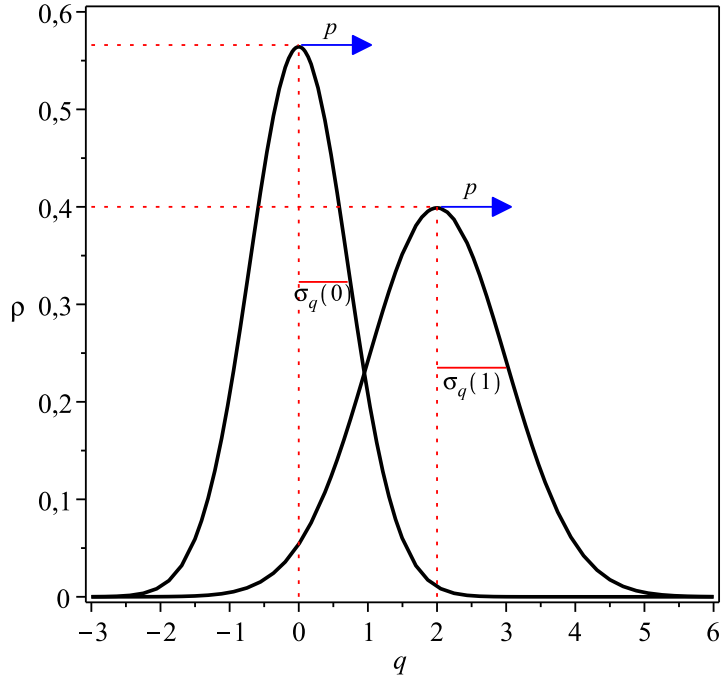}
\end{center}
\caption{Evolution of the probability densities.}
\label{fig1}
\end{figure}

Let us compare the CS (\ref{d14}) with the plane waves
\begin{equation}
\Phi_{p}\left(  q,\tau\right)  =\frac{1}{\sqrt{2\pi}}\exp\left[  i\left(
pq-\frac{p^{2}}{2}\tau\right)  \right]  .\label{d16}%
\end{equation}
Both sets of functions are solutions of the Schrödinger equation for the free
particle. The CS do belong to $L^{2}(\mathbb{R})$, whereas the plane waves do
not. A plane wave propagates with the phase velocity that is $p/2$. The CS set
generates the probability density that propagates exactly with the particle
velocity $p$. We can say that CS (\ref{d14}) represent wave packets that allow
establishing a natural connection between the classical and quantum
description of free particles. Depending of parameters of the CS, some of them
can be treated as semiclassical states of free particles, some cannot, because
they describe pure quantum states (see bellow).

\section{Semiclassical CS of free particle}

To consider the question which CS\ can be treated as representing a
semiclassical particle motion, we have to return to the initial dimensional
variables $x$ and $t$ (\ref{b2}) and to the initial wave function $\Psi\left(
x,t\right)  $ written in these variables (\ref{b4}). Taking into account that%
\begin{align}
&  x\left(  t\right)  =lq\left(  \tau\right)  =x_{0}+\frac{p_{x}}%
{m}t,\ \ p=\frac{l}{\hbar}p_{x},\nonumber\\
&  \sigma_{x}\left(  0\right)  =l\sigma_{q}\left(  0\right)  =l\sigma
=\sigma_{x},\ \sigma_{x}^{2}\left(  t\right)  =\sigma_{x}^{2}+\frac{\hbar^{2}%
}{4m^{2}\sigma_{x}^{2}}t^{2}, \label{d18a}%
\end{align}
we obtain%
\begin{align}
&  \Psi\left(  x,t\right)  =\frac{1}{\sqrt{\left(  \sigma_{x}+\frac{i\hbar
}{2m\sigma_{x}}t\right)  \sqrt{2\pi}}}\exp\left\{  \frac{i}{\hbar}\left(
p_{x}x-\frac{p_{x}^{2}}{2m}t\right)  -\frac{\left[  x-x\left(  t\right)
\right]  ^{2}}{4\left(  \sigma_{x}^{2}+\frac{\hbar}{2m}it\right)  }\right\}
,\nonumber\\
&  \rho\left(  x,t\right)  =\left\vert \Psi\left(  x,t\right)  \right\vert
^{2}=\frac{1}{\sqrt{\left(  \sigma_{x}^{2}+\frac{\hbar^{2}}{4m^{2}\sigma
_{x}^{2}}t^{2}\right)  2\pi}}\exp\left\{  -\frac{1}{2}\frac{\left[  x-x\left(
t\right)  \right]  ^{2}}{\sigma_{x}^{2}+\frac{\hbar^{2}}{4m^{2}\sigma_{x}^{2}%
}t^{2}}\right\}  . \label{d20}%
\end{align}

Semiclassical motion implies that the form of the distribution (\ref{d20})
changes slowly with time $t$ in a certain sense. This form changes due to the
change in the quantity $\frac{\hbar^{2}}{4m^{2}\sigma_{x}^{2}}t^{2}$ with
time, which is responsible for the change of $\sigma_{x}^{2}\left(  t\right)
$ (see eq. (\ref{d18a})). We suppose that in case of semiclassical motion,
this quantity is much less than the square of the distance that the particle
travel in the same time. We then have the inequality
\begin{equation}
\frac{\hbar^{2}}{4m^{2}\sigma_{x}^{2}}t^{2}\ll\left(  \frac{p_{x}}{m}t\right)
^{2}\Longrightarrow p_{x}\gg\frac{\hbar}{2\sigma_{x}}\sim\upsilon\gg
\frac{\hbar}{2m\sigma_{x}},\label{d21a}%
\end{equation}
which can be rewritten in another form:%
\begin{equation}
\lambda\ll4\pi\sigma_{x},\ \ \lambda=\frac{2\pi\hbar}{p_{x}},\label{d21b}%
\end{equation}
where $\lambda$ is the Compton wavelength of the particle. Hence, the CS of a
free particle can be considered semiclassical states if the Compton wavelength
of the particle is much less than the coordinate standard deviation
$\sigma_{x}$ at the initial instante. It is known that in a cyclotron,
nonrelativistic electrons are moving with velocities $\upsilon\simeq
10^{3}\frac{\mathrm{m}}{\mathrm{s}}$. Then, according to eq. (\ref{d21a}), CS
of such electrons with $2\sigma_{x}\simeq10^{-7}\mathrm{m}$ can be treated as
semiclassical states.

It should be noted that similar criteria of the semiclassicality were used in
theory of potential scattering \cite{Alfaro} and for classifying CS in a
magnetic-solenoid field \cite{272}.

\section{Some concluding remarks}

In this article, we have studied different types of generalized CS of a free
massive nonrelativistic particle and established properties of these states
such as the completeness relations, the minimization of uncertainty relations,
and the evolution of the corresponding probability densities in time. Among
all these types of generalized CS, families of states are naturally
distinguished which we suggest identifying with the CS of a free massive
nonrelativistic particle. These CS families are parameterized by one
real-valued parameter, the coordinate standard deviation sq at the initial
time instant. The CS from a family with a given sq form a complete system of
functions and are labeled by a complex-valued quantum number z, which is in a
one-to-one correspondence with the initial data of the corresponding
trajectory of the coordinate mean value. CS minimize the Robertson-Schrodinger
uncertainty relation at all time instants and the Heisenberg uncertainty
relations at the initial instant. The smaller the coordinate standard
deviation $\sigma_{q}\left(  \tau\right)  $ at the initial instant is, the
faster $\tau$ grows with time at an arbitrary instant. At any time instant
$\tau$, the probability density corresponding to free-particle CS is given by
Gaussian distributions with standard deviations $\sigma_{q}\left(
\tau\right)  $. The coordinate mean value propagates along the classical
trajectory with the mean particle velocity. The probability density maximum
propagates with the same velocity. The constructed CS are wave packets that
are solutions of the Schrödinger equation for a free particle. They belong to
the Hilbert space $L^{2}\left(
%TCIMACRO{\U{211d} }%
%BeginExpansion
\mathbb{R}
%EndExpansion
\right)  $, whereas plain waves do not belong to this space. The CS allow
establishing a natural relation between the classical and quantum descriptions
of free particles. Depending on the parameters of the CS, some of them can be
considered semiclassical states of free particles, and some of them cannot,
inasmuch as the latter are purely quantum states. We provide arguments in
favor of the fact that free-particle CS can be considered semiclassical states
when the Compton wavelength is much less than the standard coordinate
deviation at the initial instant. The suggested CS can be apparently
identified with the asymptotic free states in nonrelativistic quantum
scattering theory.

We believe it is useful for a lecture course in quantum mechanics to complete
the description of quantum motion of a free particle with the free-particle CS
as an example of exact wave packets, which, under certain conditions, admit a
semiclassical description of such a particle, and which allow illustrating a
large number of general principles of quantum mechanics, such as the
minimization of uncertainty relations. The acquaintance of an audience with
free-particle CS would naturally make it easier to understand the CS of an
oscillator and other quantum systems.%
\[
\]

{\Large Acknowledgement }D. Gitman thanks CNPq and FAPESP for permanent
support, the work is partially supported by the Tomsk State University
Competitiveness Improvment Program. A. S. Pereira thanks FAPESP for support.
Authors are grateful to I. Tyutin and V. Man'ko for useful discussions.

\end{document}